\documentclass[aps,prl,twocolumn,superscriptaddress,groupedaddress]{revtex4}
\usepackage{graphicx,epstopdf} 
\usepackage{amsmath}
\usepackage{color}
\usepackage{float}
\usepackage{soul}
\usepackage{booktabs}
\usepackage{multirow}

\newcommand{\be}{\begin{equation}}
\newcommand{\ee}{\end{equation}}
\newcommand{\bea}{\begin{eqnarray}}
\newcommand{\eea}{\end{eqnarray}}
\newcommand{\bse}{\begin{subequations}}
	\newcommand{\ese}{\end{subequations}}

\usepackage{multirow}

\begin{document}
\title{ Pressure controlled trimerization for switching of anomalous Hall effect in triangular antiferromagnet Mn$_3$Sn }
\author{Charanpreet Singh}
\affiliation{School of Physical Sciences, National Institute of Science Education and Research, HBNI, Jatni-752050, India}
\author{Vikram Singh}
\affiliation{School of Physics, Indian Institute of Science Education and Research Thiruvananthapuram, Kerala-695551, India}
\author{Gyandeep Pradhan}
\affiliation{School of Physical Sciences, National Institute of Science Education and Research, HBNI, Jatni-752050, India}
\author{Velaga Srihari}
\affiliation{High Pressure and Synchrotron Radiation Physics Division, Bhabha Atomic Research Centre, Mumbai 400085, India}
\author{Himanshu Kumar Poswal}
\affiliation{High Pressure and Synchrotron Radiation Physics Division, Bhabha Atomic Research Centre, Mumbai 400085, India}
\author{Ramesh Nath}
\affiliation{School of Physics, Indian Institute of Science Education and Research Thiruvananthapuram, Kerala-695551, India}
\author{Ashis K. Nandy}
\email{aknandy@niser.ac.in}
\affiliation{School of Physical Sciences, National Institute of Science Education and Research, HBNI, Jatni-752050, India}
\author{Ajaya K. Nayak}
\email{ajaya@niser.ac.in}
\affiliation{School of Physical Sciences, National Institute of Science Education and Research, HBNI, Jatni-752050, India}

\date{\today}

\begin{abstract}
Here, we present a detailed theoretical and experimental study on the pressure induced switching of anomalous Hall effect (AHE) in the triangular antiferromagnetic (AFM) compound Mn$_3$Sn. Our theoretical model suggests pressure driven significant splitting of the in-plane Mn bond lengths $i.e.$ an effective trimerization, which in turn stabilizes a helical AFM ground state by modifying the inter-plane exchange parameters in the system. We experimentally demonstrate that the AHE in Mn$_3$Sn reduces from 5~$\mu\Omega$ cm at ambient pressure to zero at an applied pressure of about 1.5~GPa. Furthermore, our pressure dependent magnetization study reveals that the conventional triangular AFM ground state of Mn$_3$Sn systematically transforms into the helical AFM phase where the symmetry does not support a non-vanishing Berry curvature required for the realization of a finite AHE. The pressure dependent x-ray diffraction (XRD) study rules out any role of structural phase transition in the observed phenomenon. In addition, the temperature dependent in-plane lattice parameter at ambient pressure is found to deviate from the monotonic behavior when the system enters into the helical AFM phase, thereby, supporting the proposed impact of trimerization in controlling the AHE. We believe that the present study makes an important contribution towards understanding the stabilization mechanism of   different magnetic ground states in Mn$_3$Sn and related materials for their potential applications pertaining to AHE switching. 
\end{abstract}


\maketitle

\section{INTRODUCTION}

Effective manipulation of magnetic structure can decisively control the electrical \cite{AHEferroReview,AHEreviewNagaosa,AHEtheorynonC,AHEmn3SnNakatsuji,AHEmn3GeAjaya}, thermal \cite{THEtheory} and optical \cite{opticskerrtheory,optickerrTopo,AFopticsreviw} properties in magnetic materials. In case of electrical transport in a magnetic system, the spins of the conduction electrons exhibit a strong Hund's coupling with the underlying magnetic structure. The fictitious magnetic field arising from such an interaction can significantly modify the path of the conduction electrons, leading to the observation of anomalous Hall effect (AHE) and topological Hall effect (THE) in ferromagnets \cite{AHEferroReview}/antiferromagnets \cite{AHEtheorynonC,AHEmn3SnNakatsuji,mn3snPredeep}. A lot of effort has been put forward to study the possibility of switching these transport signals to facilitate device applications. Although the AHE in ferromagnets has a great potential for its utilization as a memory element in spintronic devices, the only viable mechanism seems to be  the reversal of underlying magnetic structure. In this regard, antiferromagnets with diverse magnetic structures are excellent candidates  to achieve a control over the AHE by tuning the magnetic ground state. The fact that antiferromagnets are extremely stable against external perturbations makes them attractive candidates for spintronics. Although electrical switching of antiferromagnets have been demonstrated recently \cite{EswitchJung,Leandro}, yet the realization of such a scenario in real devices remains afar.

\begin{figure*} [tb!]
	\includegraphics[width=17cm,clip]{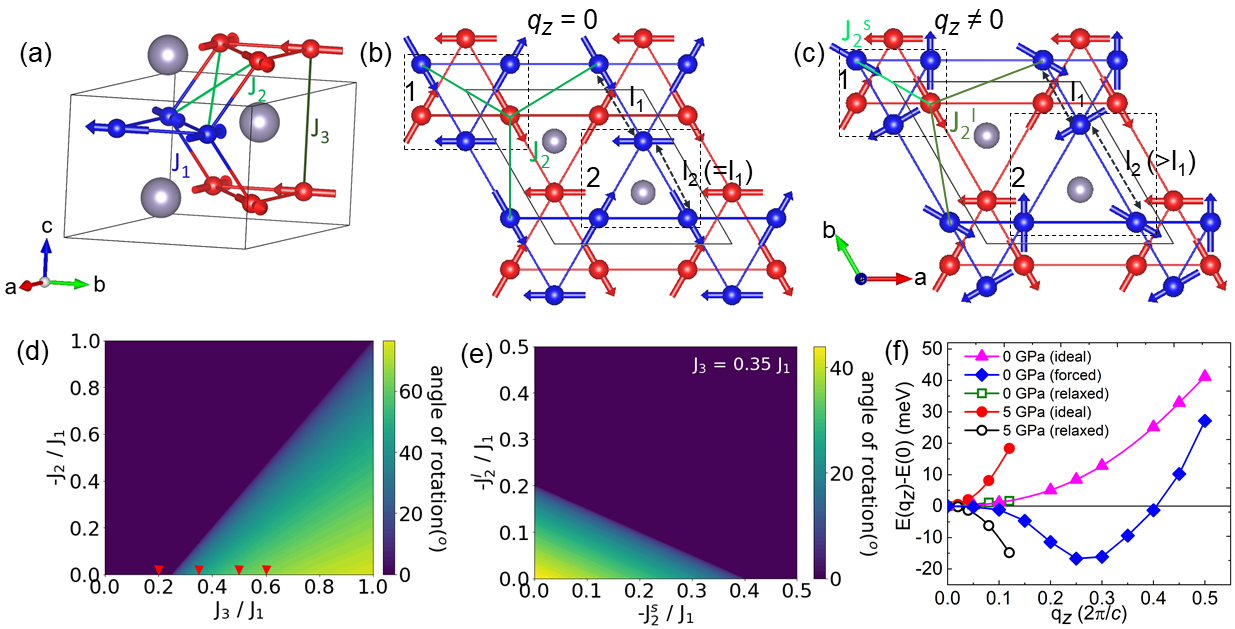}
	\caption{\label{fig1}(Color online) (a) Unit cell for hexagonal Mn$_3$Sn and stacking of three triangular layers (Kagome layers) of Mn moments with $120^{\circ}$ AFM spin structure. Red and blue balls represent different layers of Mn atoms, whereas, the Sn atoms are shown in gray balls. $J_1$, $J_2 $, and $ J_3 $ represent first three out of plane (interlayer) neighboring exchange interactions between the Mn moments. (b) Top view of Mn$_3$Sn where two ideal Kagome sublattice of Mn atoms are stacked. The inplane bond lengths $l_1$ and $l_2$  of the equilateral triangles shown in dashed boxes $\lq$1' and $\lq$2' and here $\delta$ = $\frac{l_2 - l_1}{l_2 + l_1}$ = 0. (c) Top view of the distorted (relaxed) Kagome sublattice stacking in Mn$_3$Sn with two different inplane bond lengths ($l_2 > l_1$) for the equilateral triangles as shown in dashed boxes $\lq$1' and $\lq$2' with $\delta$ $ \neq $ 0. Splitting of $J_2$ to $J_2^l$ and $J_2^s$ is shown in green lines.  (d) Phase diagram in the $J_2$ - $J_3$ plane for ideal Kagome layer stacking. $J_2$ and $J_3$ are measured in the unit of the magnitude of J$_1$. The color code indicates the angle of rotation of Mn triangle spins between two layers due to the helical modulation. The deep blue region indicates the $120^{\circ}$ triangular AFM state. (e) Phase diagram in the $J_2^l$ - $J_2^s$ plane where the nonzero $\delta$ value results in splitting of $J_2$ to $J_2^l$ and $J_2^s$.  The values marked by the down triangles on the $J_3$ axis in (d) are used for the phase diagrams shown in Fig. \ref{S2} (Appendix A). (f) \textit{Ab initio} calculated spin spiral energy as a function of the spin spiral  vector $\textbf{q}$ = $(0,0,q_z)$ where $120^{\circ}$ triangular AFM state corresponds to $\textbf{q}$ = 0. Energies of the spin spiral state are calculated with respect to the $120^{\circ}$ AFM state at zero pressure for $\delta $= 0 (filled triangles), $\delta $= 2.4 (open squares), and $\delta $= 12.0 (filled diamonds). The data for $\delta $= 0 at the zero pressure (filled triangles) are divided by a factor of ten for better visibility and comparison in the plot. The $\textbf q$ dependence of the spiral energy at pressure 5~GPa in case of ideal ($\delta$ = 0) and relaxed ($\delta$ = 10.34) Kagome structure are also shown with filled and open circles, respectively. The symbols are calculated data points and the solid lines are to guide the eye. The negative slope in the dispersion curves indicates the helical modulation of the $120^{\circ}$ triangular AFM state along $z$.}
\end{figure*}

In particular, the non-collinear antiferromagnets are extensively pursued as their magnetic states can be easily  manipulated by spin transfer torque related mechanisms.  Recently, the hexagonal crystal structure based triangular antiferromagnets Mn$ _3 $Sn and Mn$ _3 $Ge have received a special attention due to their various anomalous transport properties, both in bulk systems \cite{AHEmn3SnNakatsuji,AHEmn3GeAjaya,teraAnomaly,invSpinMn3sn} and  thin films \cite{mn3snfilm1,mn3snfilm2,mn3snfilm3} . In the hexagonal structure of Mn$ _3 $Sn, the Mn atoms form a kagome-type lattice with Sn atoms residing at the center of the hexagon. The geometrical frustration arising due to the AFM coupling of Mn moments and the kagome type of arrangement of Mn atoms incite a $120^{\circ}$ magnetic ground state \cite{mn3snNutron}. In this structure, two layers of Mn atoms in a unit cell together form an inverse triangular spin structure as depicted in Fig.~\ref{fig1}(a). The topological Weyl semi-metallic nature of these materials has been demonstrated by theoretical calculations \cite{theoryWeylmn3snge_Kubler,theoryWeylmn3snge_BYan}, as well as experimental studies \cite{expWeylmn3sn}.  The presence of Weyl nodes near the Fermi level is believed to be the source of large AHE in these materials \cite{AHEweylTheory,expWeylmn3sn,theoryWeylmn3snge_BYan}. These Weyl nodes occur in pairs with opposite chirality and act as magnetic monopoles in the momentum space. An electron crossing between two Weyl nodes of opposite chirality picks up an extra Berry phase, and consequently a large AHE has been observed in case of Mn$_3$Sn and Mn$_3$Ge. Since the topological Weyl nature, and hence the AHE of these materials is critically inherited from the special kind of magnetic structure, a precise control over the magnetic state is essential for any possible switching of the AHE.

 Mn$ _3 $Sn exhibits distinct AFM ground states in different temperature regimes depending upon the Mn concentration in the system \cite{diffMncomposition}. While most of the samples show an in-plane triangular AFM ordering at room temperature, the presence of a slightly canted triangular AFM structure has also been found at very low temperatures \cite{Fengcaxis,Tomicaxis}. The recent observation of large THE below 50~K in Mn$_3$Sn corroborates the existence of noncoplanar magnetic structure at low temperatures \cite{mn3snPredeep}. While the samples used by Kuroda \textit{et al.} \cite{expWeylmn3sn} display an additional magnetic transition below 50~K, the presence of a sharp transition around 240~K related to the helical magnetic ordering has also been observed in other systems \cite{mn3sndiffphase}. Irrespective of the nature of the low temperature transitions, all the samples display a large AHE at room temperature. However, the AHE ceases to zero below 240~K when the system enters into a spin spiral (SS) state \cite{mn3sndiffphase}. Using $ab~initio$ calculations and Monte Carlo simulations, Park~\textit{et al.} have demonstrated the presence of an incommensurate helical modulation of the triangular spin structure that propagates along the $c$-direction \cite{mn3qm}. Interestingly, no Weyl nodes were found in the electronic band structure for the SS magnetic state, explaining the absence of AHE in the helical magnetic phase \cite{mn3qm}. Therefore, it is important to investigate the correlation between the two observed magnetic ground states of Mn$_3$Sn to explore the possibility of controlling the transition between them. Motivated by the aforementioned discussion, herein, we utilize hydrostatic pressure as a control parameter to probe the switching mechanism of AHE in case of the non-collinear antiferromagnetic Mn$_3$Sn.


\section{Methods}
The density functional theory (DFT) calculations were employed to study the magnetic properties of the present system. 
The spin-polarized calculations was carried out within the projector augmented wave (PAW) method \cite{paw} as implemented in Vienna Ab initio Simulation Package (VASP) \cite{V1,V2,V3,V4}. 
The generalized gradient approximation (GGA) was used as the exchange-correlation potential in the form of Perdew-Burke-Enzerhof (PBE) \cite{pbe}. 
For the Brillouin-Zone (BZ) integration, a $\Gamma$-centered $k$-point grid of 8$\times$8$\times$9 was used with a plane-wave cut-off energy of 500 eV. 
Note that the noncollinear version of the code was utilized for determining the conventional magnetic ground state in this type of hexagonal system. 
The experimentally obtained lattice constants were kept fixed in our calculations where the internal coordinates were relaxed till the forces become less than 1~meV/{\AA}. 
Further, the spin-spiral calculations were performed using the full potential linearized augmented plane wave (FLAPW) method implementing the noncollinear version of the FLEUR code \cite{fleur1}. 
The energy of the SS state characterized by wave vector {\textbf q} was calculated using the generalized Bloch theorem \cite{fleur2}. 
We have also employed PBE exchange-correlation functional as the GGA approximation. 
We have considered 16$\times$16$\times$18 mesh for $k$-points in the full-BZ and a cutoff of $K_{max}$ = 4.1 $a.u.^{-1}$ to expand the LAPW basis functions.
The {\textbf q} =0 state was considered as a perfect $120 ^{\circ}$ triangular spin configuration on each Kagome sublattice.

Polycrystalline ingots of Mn$ _{3+x} $Sn$ _{1-x} $ with $ x= $ 0.05, 0.04, and 0.03 were prepared by arc melting stoichiometric amount of high pure Mn and Sn metals under argon atmosphere.  The samples were melted multiple times to ensure a better homogeneity. The as-prepared ingots were annealed for eight days in evacuated quartz tubes at a temperature of 1073~K, followed by slow cooling to room temperature.
Room temperature x-ray powder diffraction measurements were performed using a Rigaku SmartLab x-ray diffractometer with a Cu-K$\alpha$ source. Low temperature XRD measurements were carried out by utilizing a low temperature attachment (Oxford Pheonix)  to the x-ray diffractometer (PANalytical). To investigate the crystal structure with applied pressure, XRD measurements were performed using a diamond anvil cell at the beamline BL-11, INDUS-II synchrotron radiation source (Raja Ramanna Centre for Advanced Technology, Indore, India). Magnetic and transport measurements were carried out using a Quantum Design (QD) SQUID magnetometer and QD Physical Property Measurement System (PPMS), respectively.  Pressure dependent magnetic and transport studies were conducted using respective pressure cells provided by Quantum Design.

\section{RESULTS AND DISCUSSION}
It is very important to understand the underlying magnetic states which govern the control and switching of AHE in Mn$_3$Sn. The primary contributions to the total energy of the magnetic state can be narrowed down to the exchange interactions, Dzyaloshinskii-Moriya interaction (DMI), and anisotropy energy. Since the inverse triangular spin structure in Mn$_3$Sn is mainly stabilized by geometrical frustration, the role of DMI to influence the long-range magnetic-order in the present centrosymmetric system is extremely limited \cite{moriyaDMI}, except breaking the degeneracy of the sense of rotation. The contribution from the anisotropy energy can also be safely ruled out for the helical modulation of an in-plane triangular spin structure along $z$-axis. Thus, the competing out of plane exchange interactions primarily determine the helical modulation in the system. To understand the origin of the helically modulated magnetic ground state, we first analyze the role of different exchange parameters within classical Heisenberg model,  $H = \sum_{i>j} J_{ij} \hat{n}_i \cdot \hat{n}_j$, where the exchange constants $J_{ij}$s determine the strength and nature of interactions between Mn moments pointing along the unit vector $\hat{n}$.

In the hexagonal Mn$_3$Sn structure each Mn atom in the Kagome plane  has four nearest neighbors, where two equilateral triangles are connected by a common Mn atom [see the lattice with red/blue Mn balls in Fig \ref{fig1}(b)]. However, it has been reported that  Mn$_3$Sn forms a distorted Kagome lattice where one of the  Mn triangle shrinks while the other expands as shown in Fig.~\ref{fig1}(c) \cite{ANEmn3sn}. Here, we first consider the case of an ideal Kagome sublattice structure and try to find out the role of frustration in the exchange parameters in stabilizing the SS ground state. For this purpose, we made a reasonable assumption that the in-plane exchange constants do not play any role in stabilizing the helical order along the $z$-direction. Therefore, we fixed the in-plane magnetic structure to a perfect $120 ^{\circ} $ triangular AFM configuration. The first three inter-plane nearest-neighbor exchange parameters, $J_1$, $J_2$, and $J_3$ as labeled in Fig.~\ref{fig1}(a), are taken into consideration in order to study the transition from noncollinear  AFM state to  helically modulated magnetic state. In the inelastic neutron scattering experiment, $J_1$ and $J_3$ were reported to be AFM in nature while $J_2$ was found to be ferromagnetic (FM) \cite{mn3qm}.  The energy of the magnetic state  per Mn atom is now translated into the following classical exchange Hamiltonian,
\begin{multline} \label{Hemil}
H = 2J_1\cos(\frac{2\pi}{3}+\theta) +2J_1\cos(\frac{2\pi}{3}-\theta)+6J_2\cos \theta \\ 
+2J_3\cos(2\theta).
\end{multline}
Here, $\theta$ is the measure of helical modulation {\textit i.e.} the angle of rotation between two successive layers of spin-triangles connected with the SS vector {\textbf q} and $\frac{2\pi}{3}$ in $J_1$ is the initial phase between two consecutive Kagome sublattices.
The value of $\theta$ varies from 0 to $180 ^{\circ}$ and the corresponding lowest energy state is the ground state in the $J_2$-$J_3$ plane.  Here,  both $J_2 (\le 0)$ and $J_3 (\ge 0)$ are relative quantity with respect to the magnitude of $J_1$. The results presented in Fig.~\ref{fig1}(d) show a sharp boundary between the noncollinear AFM ordering with and without the helical modulation along $z$. We find that the competing $J_2$ and $J_3$ drives the system into the helical SS state, indicating that the exchange frustration is the key route for achieving the spiral modulation along $z$.  It is important to mention here that the SS ground states stabilized by competing exchange interactions have also been found in hexagonal \cite{femono} and square \cite{NandyPRL} lattices.

The ideal Kagome sublattice in Mn$_3$Sn is expected to be distorted in two unequal equilateral triangles as discussed above. Therefore, we define an effective trimerization parameter, $\delta$ as $\frac{l_2 - l_1}{l_2 + l_1}$, where $l_1$ and $l_2$ are the bond lengths corresponding to two equilateral triangles formed by Mn atoms in the Kagome sublattice [see dashed square $\lq$1' and $\lq$2' in Fig.~\ref{fig1}(b-c)]. One can consider $\delta$ as the effective trimerization parameter where the inplane exchange coupling within two unequal triangles can be changed significantly \cite{mn3qm,gd3trimer}. Here we note that $\delta=0$ defines the ideal Kagome lattice. 
Since a nonzero value of $\delta$ causes bond length mismatch related to the exchange constant $J_2$, corresponding large and small out of plane exchange constants $J_2^l$ and $J_2^s$ are expected to change their strength significantly and even their sign too. We, therefore, extend our analyses by constructing phase diagrams in the $J_2^l$ - $J_2^s$ plane for several values of $J_3$ marked by arrows in Fig.~\ref{fig1}(d). Figure.~\ref{fig1}(e) shows the $J_2^l$ - $J_2^s$ phase diagram for  $J_3 =0.35$. Similar phase diagrams for $J_3 =0.2$ (inside non-helical AFM phase) and $J_3 =0.6$ (inside the helical AFM phase) with all possible $\lq$sign' combinations of $J_2^l$ and $J_2^s$ are shown in Fig. \ref{S2} (Appendix A). It is evident that the area of helical magnetic state increases with increasing $J_3$ value {\textit i.e.} the $\frac{J_3}{J_1}$ ratio. It is therefore remarkable that  the control in $\delta$ values plays a crucial role in determining the magnetic state of the hexagonal systems like Mn$_3$Sn.


\begin{table}[ht]
	\centering
	\caption{\label{tab1} $\delta$, the measure of effective trimerization at different pressure values in case of ideal and relaxed structures along with Mn Wyckoff position.}
	\begin{tabular}[t]{lcc}
		\hline
		Pressure (GPa)  &  $x$ (6c: $x$,$2x$,0.25)   &    $\delta$ (10$^{-2}$)\\
		\hline
		0.0 (ideal)         &      0.1666         &     0.0    \\
		5.23 (ideal)        &      0.1666        &      0.0   \\
		0.0 (forced)      &      0.1466         &     12.0  \\
		0.0 (relaxed)     &      0.1626         &     2.4     \\
		2.75 (relaxed)   &      0.1605         &      3.7     \\
		5.23 (relaxed)   &      0.1493         &     10.34 \\
		\hline
	\end{tabular}
\end{table}%


The conventional $120 ^{\circ}$ triangular AFM ground state can undergo a phase transition into the helical SS state  by means of effective trimerization, which may bring significant changes in $J$'s due to the changes in bond lengths. 
Hence, the effect of  pressure $P$ in the hexagonal Mn$_3$Sn system has been carried out within {\textit{ab initio}} electronic structure calculations where hydrostatic pressure $P$ is used as a parameter. The last column in Table~\ref{tab1} lists the $\delta$ values  for various pressures. The corresponding Wyckoff position of Mn atom at the middle column determines the Kagome sublattice structures. Interestingly, the application of pressure, chemical as well as hydrostatic, results in a nonzero $\delta$ value that keeps on increasing with increasing pressure. Figure~\ref{fig1}(f) summarizes the calculated energy corresponding to the noncollinear magnetic SS states characterized by $\textbf q$. In order to understand the role of pressure on $\delta$, and hence the magnetic ground state, we have taken $120 ^{\circ}$ inverse triangular AFM state as the initial state corresponding to $\textbf q$ = 0. Subsequently, the triangular spin state is helically modulated for $q_z \ne$ 0, as can be seen by comparing the spin configurations in Fig.~\ref{fig1}(b) and (c). The ideal Kagome sublattice stacked Mn$_3$Sn shows noncollinear $120 ^{\circ}$ inverse triangular AFM structure as the ground state up to a hydrostatic pressure $P= 5.23$ GPa. The SS dispersion does not show any significant changes like softening or hardening near the ground state (see Fig. \ref{S3} (Appendix A)). However, the forced structure with $\delta$ = 0.12 at ambient pressure ($P$ = 0) shows a SS ground state with a period of about  16.3 {\AA}. Such a helically modulated spin structure is indeed stabilized by the effective trimerization of the system which is fostered by pressure.

To interpret our calculations, we map the results onto a classical Heisenberg model which allows us to determine the first three/four inter sublattice exchange constants as discussed earlier. A good fit is obtained by using $J_1$, $J_2$, and $J_3$ for the ambient pressure ideal structure where the nature of exchange constants is similar to the experimental results \cite{mn3qm}. In case of forced structure at ambient pressure, the four exchange constants which include $J^s_2$ and $J^l_2$ give good fitting to the results with SS ground state. Details of the calculation is provided in Appendix A. We find that the ratio of $\frac{J_3}{J_1}$ does not change significantly when we move from the ideal to the forced trimerization case. Interestingly, $J^s_2$ changes sign to become antiferromagnetic while $J^l_2$ remains ferromagnetic.   For the ambient pressure ideal structure, the relative values of $J_2$ and $J_3$ measured in the unit of $J_1$ are about 0.445 and 0.340, respectively. The corresponding point lies in the dark blue region in Fig.~\ref{fig1}(d). In case of the forced trimer system, the  $J_3$ value is about 0.336, which is slightly smaller than the value used for constructing Fig.~\ref{fig1}(e). This is indeed remarkable as our calculated $J^s_2$ and $J^l_2$ (in the unit of $J_1$) lie in the region of SS phase in Fig. 1(e) [see table \ref{tab2} (Appendix A)]. Now, the relaxed structure at ambient pressure changes $\delta$ value from 0 (ideal) to 0.024 due to the chemical pressure effect. The SS energy dispersion calculated up to $\textbf q$ = (0,0,0.125) shows significant softening, possibly due to the effective trimerization. The relaxed structure at $ P $= 5.23 GPa shows significantly large $\delta$ value of about 0.10, which further establishes the role of pressure in stabilizing the helical ground state. Interestingly, the dispersion shows negative slope near $\textbf q$ = 0, indicating the ground state at finite $(0,0,q_z)$ value. Hence, the pressure induced effective trimerization in the Kagome sublattices modifies the frustrated exchange constants which ultimately determine the magnetic ground state. 


\begin{figure}	
	\includegraphics[angle=0,width=8.5cm,clip]{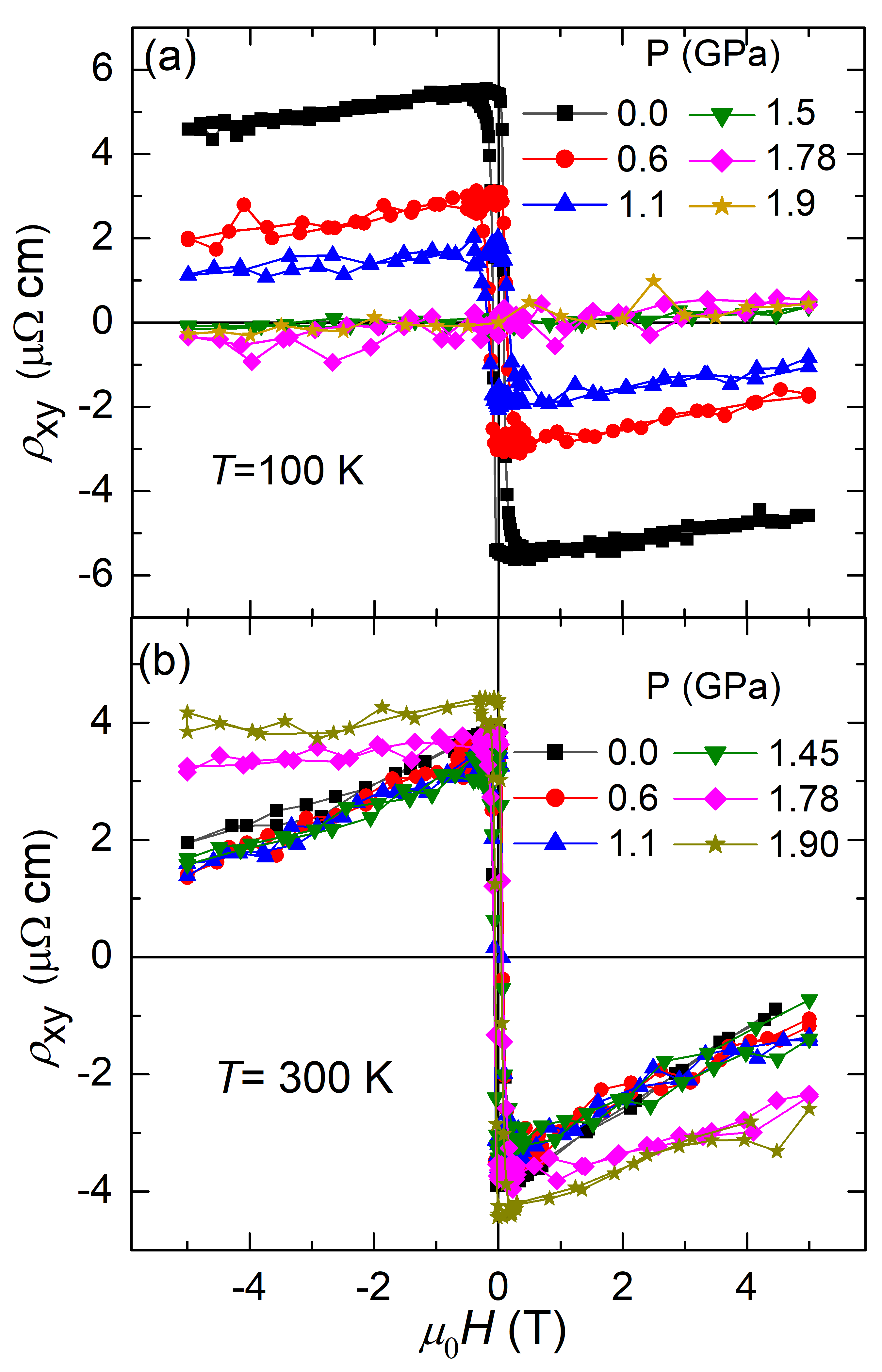}
	\caption{\label{Fig2}(Color online) Field dependence of Hall resistivity for Mn$_{3.05}$Sn$ _{0.95} $ at different hydrostatic pressures measured at (a) $T = 100$~K and (b) $T = 300$~K.}
\end{figure}

\begin{figure}	
	\includegraphics[angle=0,width=8.5cm,clip]{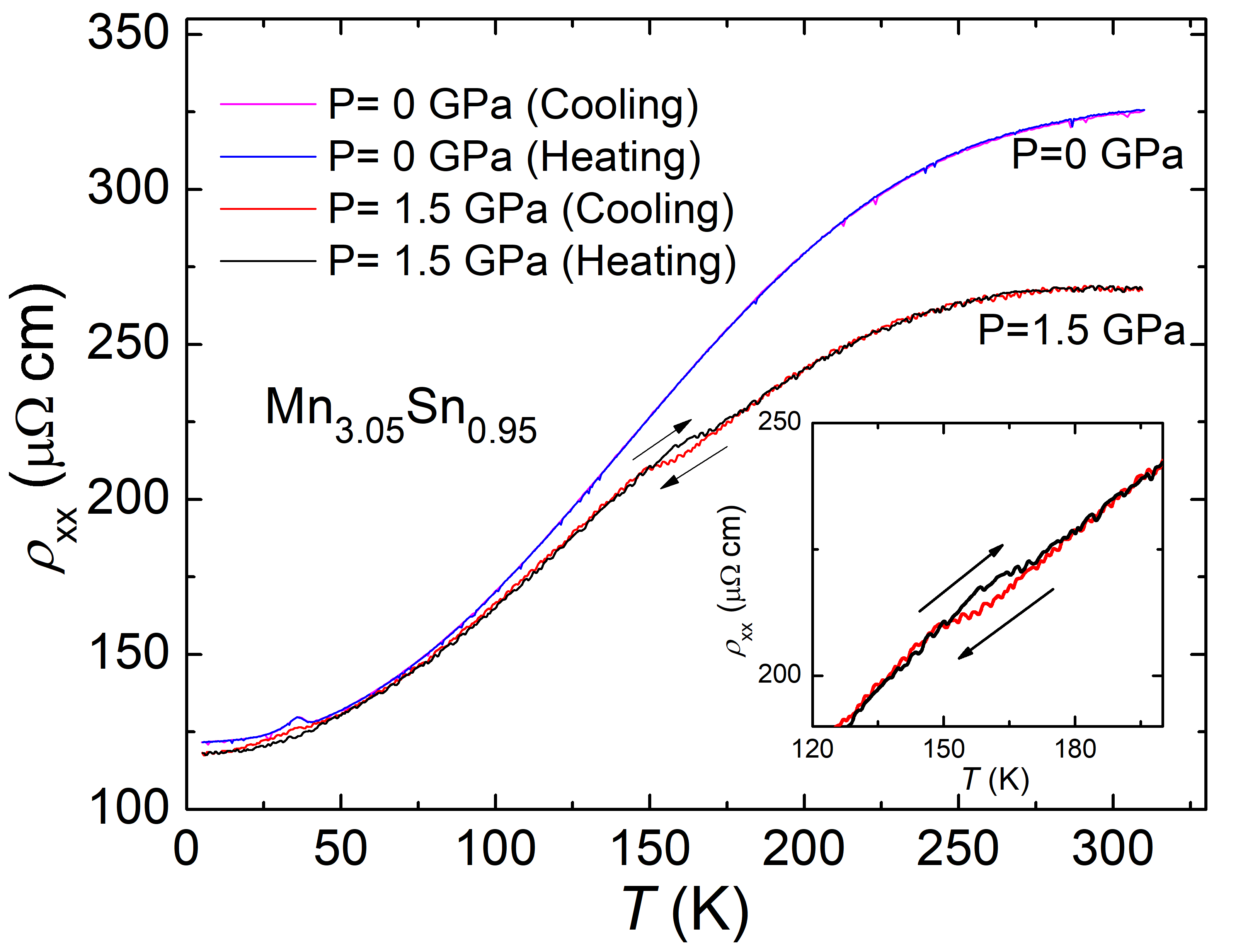}
	\caption{\label{fig3}(Color online) 	Temperature dependent zero field longitudinal resistivity [$ \rho_{xx}(T) $] measured at ambient and 1.5 GPa pressure for Mn$_{3.05}$Sn$_{0.95}$. The measurement is performed both in cooling and heating modes as marked by arrows. The inset shows an amplified view of the 1.5 GPa $ \rho_{xx}(T) $ data around the phase transition region. A small kink  at 35 K present in the zero pressure data corresponds to the instrumental artifact arising from the inconel used in the PPMS.}	
\end{figure}


Guided by our theoretical results, we set out to explore the possibility of controlling the magnetic ground state and its effect on the AHE in Mn$ _3 $Sn with the help of hydrostatic pressure as a control parameter. The structural characterization for the samples studied in the present manuscript can be found in Appendix B. Figure 2 shows the field dependence of anomalous Hall resistivity ($ \rho_{xy} $) measured at 100~K and 300~K at different applied pressures. As one can see, the $ \rho_{xy} $ value at 100~K monotonically decreases from a large value of 5~$\mu\Omega$ cm at ambient pressure to almost zero by increasing the pressure to 1.5~GPa. It is worth mentioning that the magnitude of AHE at ambient pressure matches well with the previously reported value \cite{AHEmn3SnNakatsuji}. A further  increase in pressure does not affect the anomalous component of $ \rho_{xy} $ that remains at zero. Contrary to the $ \rho_{xy} $ behavior at 100~K, the anomalous Hall resistivity at 300~K remains almost constant when the pressure is changed from 0~GPa to 1.45~GPa, before showing a slight increasing trend at higher pressures. To further support the observed pressure induced switching of AHE in the present system, we have measured temperature dependent longitudinal resistivity [$ \rho_{xx}(T) $] at ambient and 1.5 GPa pressures (Fig. 3). The ambient pressure  $ \rho_{xx}(T) $ curves measured both in cooling and heating modes exhibit a metallic type of temperature dependency without the signature of any transition, as reported before \cite{AHEmn3SnNakatsuji}. In contrast, the $ \rho_{xx}(T) $ data measured at 1.5 GPa display a transition like anomaly with the appearance of a small hysteretic behavior between the cooling and heating $ \rho_{xx}(T) $ curves at about 150 K. The first-order nature of this transition suggests the presence of a coupled magnetic and structural transition might be responsible for the pressure induced switching of AHE in the present system. 

\begin{figure}	
	\includegraphics[angle=0,width=8.5cm,clip]{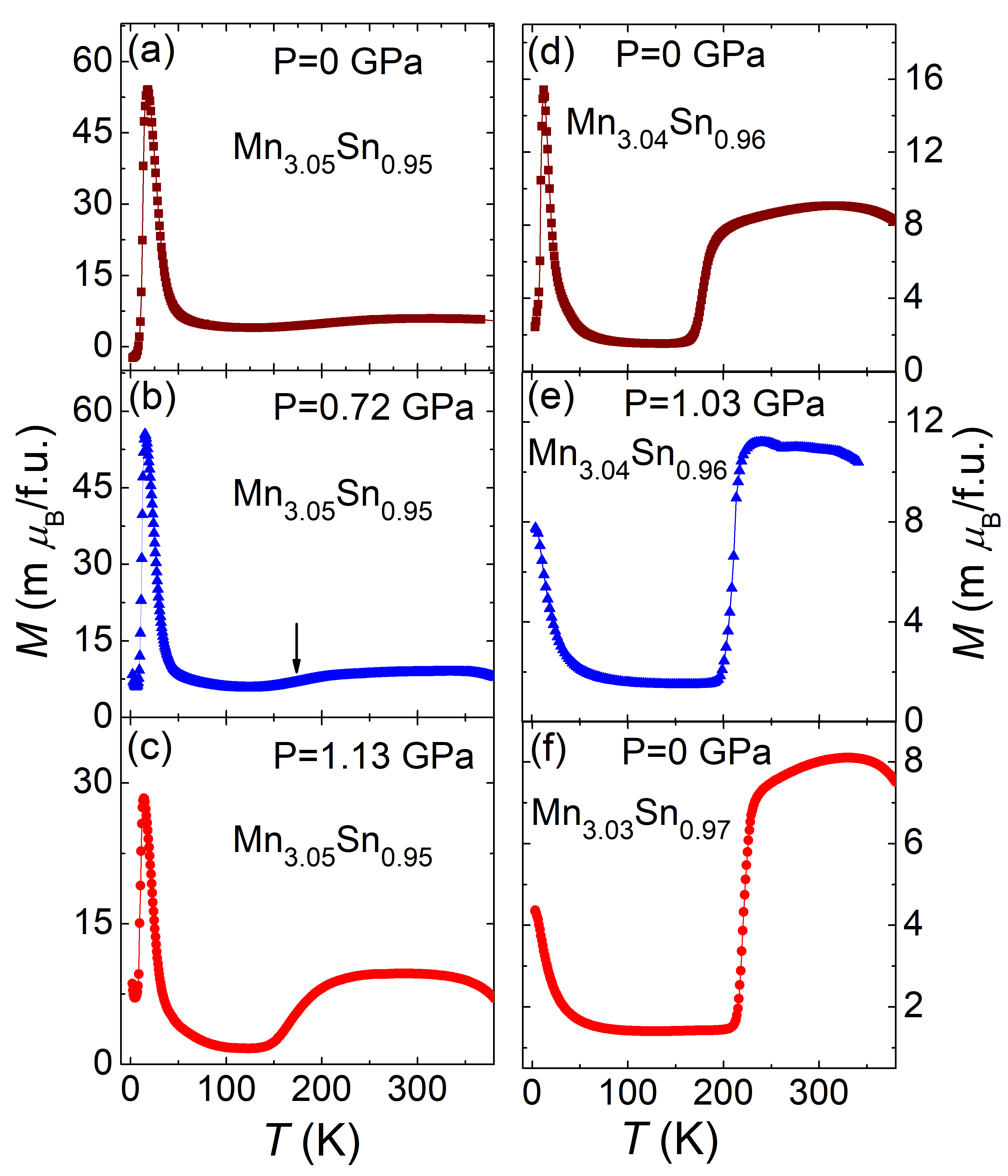}
	\caption{\label{fig4}(Color online) Temperature dependence of zero field cooled (ZFC) magnetization [$ M(T) $]  for Mn$_{3.05}$Sn$_{0.95}$ sample measured in an applied pressure of (a) $P = 0$~GPa, (b) $P = 0.72$~GPa, and (c) $P = 1.92$~GPa. $M(T)$ curves for Mn$_{3.04}$Sn$_{0.96}$ sample measured in (d) $P = 0$~GPa and (e) $P = 1.03$~GPa. (f) $ M(T) $ curve for Mn$_{3.03}$Sn$_{0.97}$ measured in ambient pressure. All the magnetization measurements were performed in an applied field of 0.1~T. }	
\end{figure}


So far  all the Hall effect measurements are carried out in the sample with chemical composition Mn$ _{3.05} $Sn$ _{0.95} $. To understand the contrasting nature of the pressure dependent AHE at 100~K and 300~K, we have carried out a thorough magnetization study in the presence of  hydrostatic pressure for three different samples with a small variation in the Mn/Sn ratio.  The temperature dependence of magnetization [$ M(T) $] curves measured at three different pressures and temperature up to 390~K are shown in Figs.~4(a)-(c). The zero pressure $ M(T) $ curve exhibits almost a temperature independent behavior down to 50~K, below which the system undergoes a transition to the canted AFM state as reported earlier \cite{Tomicaxis}. Since we mainly focus  on the switching of AHE in the triangular AFM phase, the rest of our discussion will be concentrated on the results  above 50~K. The $ M(T) $ curve measured at a pressure of 0.72~GPa develops a small step-like feature at about 150~K, marked by an arrow in Fig.~4(b). This additional transition at 150~K becomes very prominent when the pressure is further increased to 1.13~GPa [see Fig.~4(c)]. To probe the origin of this new transition with pressure we have collected $ M(T) $ data for Mn$ _{3.04} $Sn$ _{0.96} $ which consists of a slightly lower Mn concentration, as depicted in Fig.~4(d). Interestingly, the ambient pressure $ M(T) $ data for Mn$ _{3.04} $Sn$ _{0.96} $ display a similar transition as that is observed at 1.13~GPa for Mn$ _{3.05} $Sn$ _{0.95} $.  The additional transition in Mn$ _{3.04} $Sn$ _{0.96} $ shifts to about 200~K at 1.03~GPa [Fig.~4(e)]. Similar pressure dependent shifting of the low temperature transition has also been reported earlier \cite{IEEE2017}.  By further reducing the Mn concentration in Mn$ _{3.03} $Sn$ _{0.97} $, the ambient pressure transition in the $ M(T) $ appears at around 220~K [Fig.~4(f)]. Here we note that lke the $ M(T) $ data, the ambient pressure $ \rho_{xx}(T) $ data for Mn$ _{3.03} $Sn$ _{0.97} $ also display a transition around 220 K \cite{Suppl}.  As mentioned in the introduction section, Mn$ _3 $Sn  exhibits multiple magnetic phases depending on the Mn concentration. The extra transition found in case of Mn$ _{3.04} $Sn$ _{0.96} $ and Mn$ _{3.03} $Sn$ _{0.97} $ at ambient pressure corresponds to the magnetic phase transition from the high temperature triangular AFM state to the helical AFM phase at low temperatures. The application of hydrostatic pressure shifts this transition to higher temperatures. Hence the appearance of the extra transition in case of Mn$ _{3.05} $Sn$ _{0.95} $ actually represents a pressure induced magnetic phase transition from the conventional triangular AFM to the helical AFM phase. Since the helical AFM structure does not support any AHE, the observation of switching of anomalous Hall resistivity in Fig.~2(a) indeed corresponds to the magnetic phase transition in the sample.

\begin{figure}
	
	\includegraphics[angle=0,width=8.5cm,clip]{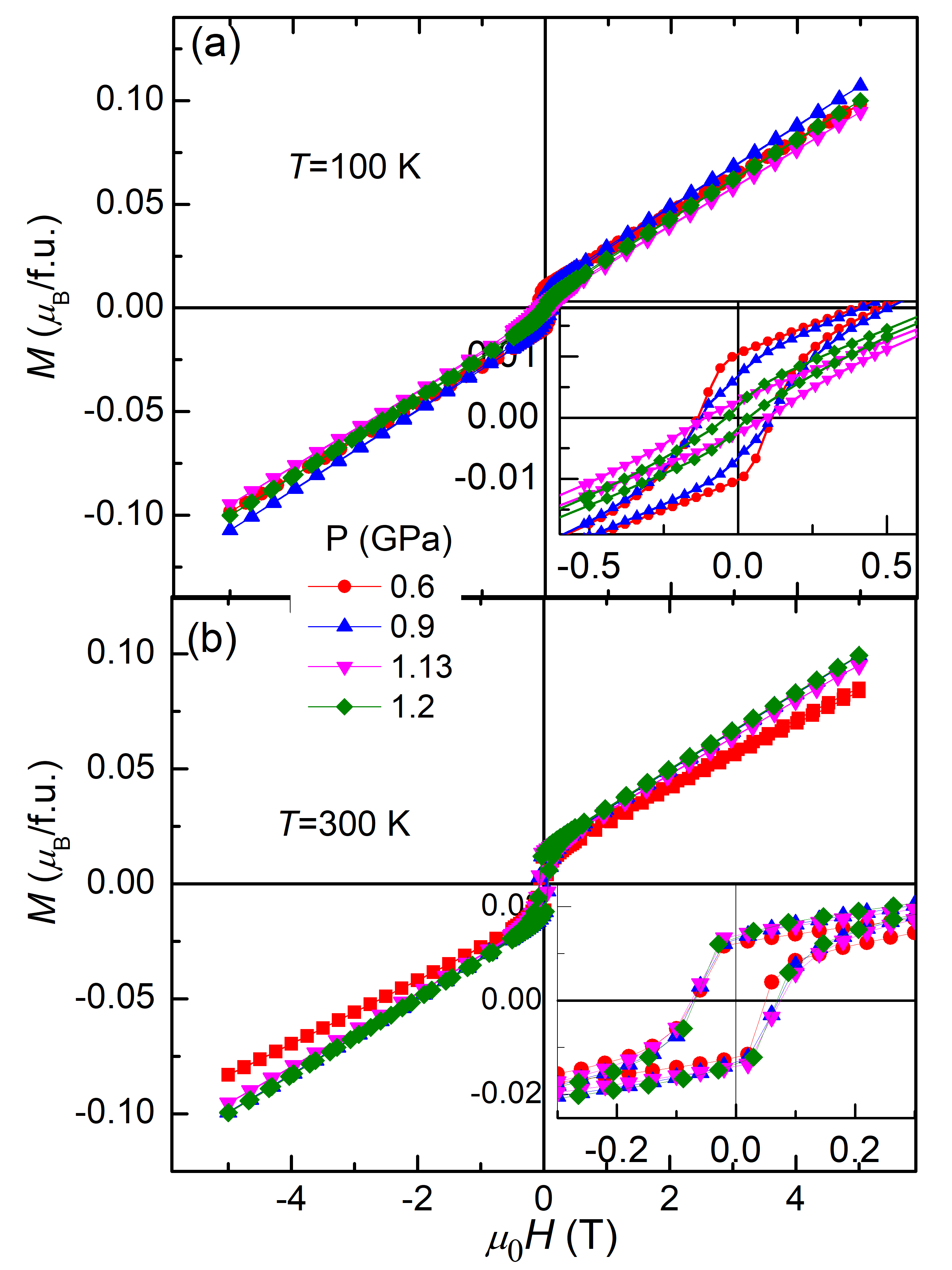}
	\caption{\label{fig5}(Color online) Field dependent isothermal magnetization loops [$ M(H) $] for Mn$_{3.05}$Sn$_{0.95}$ sample measured at different hydrostatic pressures and at two temperatures (a) $T = 100$~K and (b) $T = 300$~K. Insets: the magnified magnetization data in the low field regime.}
	
\end{figure}

\begin{figure}
	
	\includegraphics[angle=0,width=8.5cm,clip]{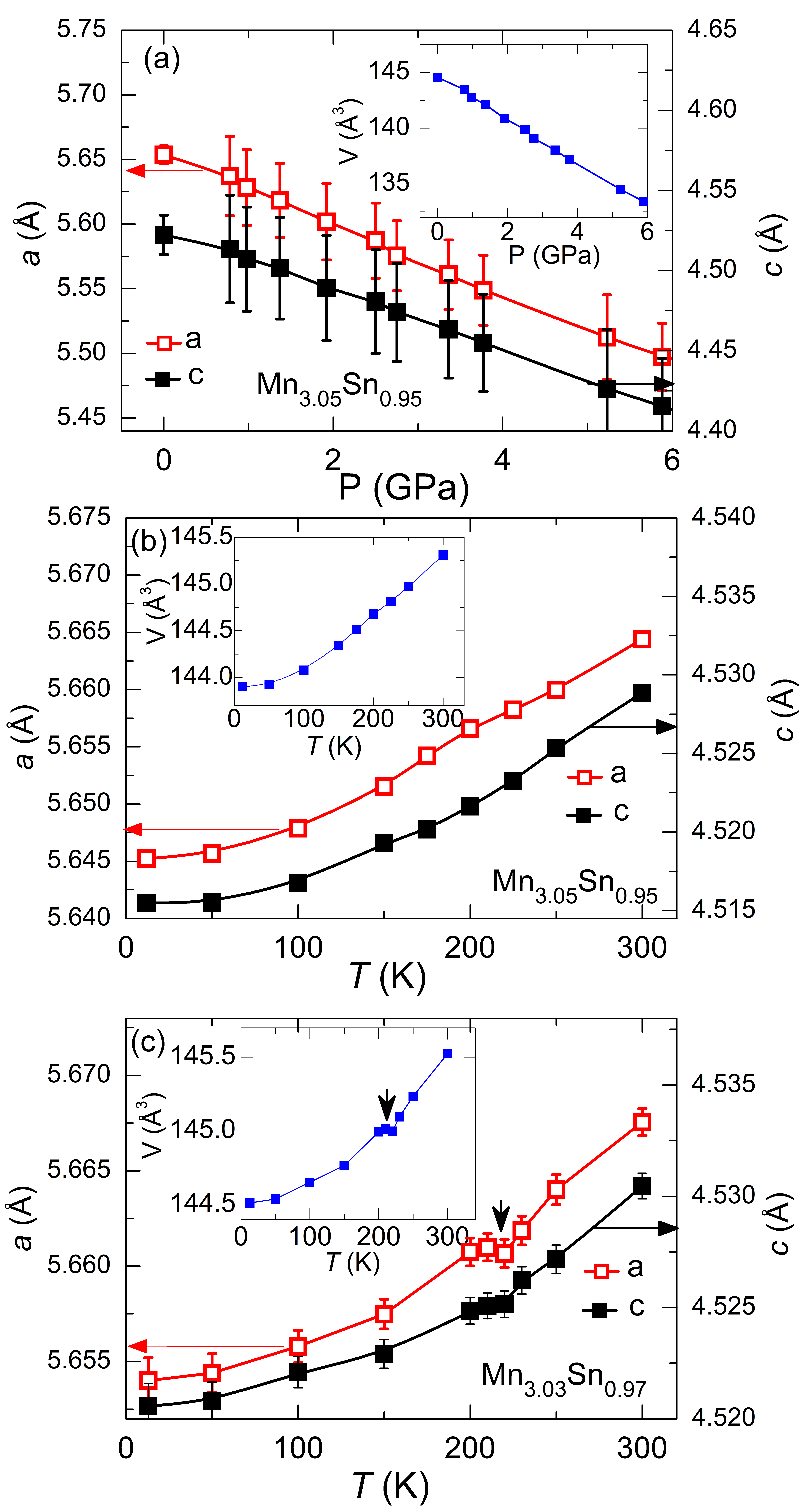}
	\caption{\label{fig6}(Color online) (a) Pressure dependence of lattice parameters $ a  $ (left ordinate) and $ c $ (right ordinate) for the hexagonal Mn$_{3.05}$Sn$_{0.95}$ calculated at room temperature. Inset: The unit cell volume as a function of hydrosatic pressure. Temperature dependence of lattice parameters $ a $ and $ c $ for (b) Mn$_{3.05}$Sn$_{0.95}$ and (c) Mn$_{3.03}$Sn$_{0.97}$ in ambient pressure. Insets of (b) and (c) present the unit cell volume as a function of temperature. The temperature at which the change of lattice parameters/unit cell volume occurs is marked by a downward arrow.}
	
\end{figure}

Our pressure dependent $ M(T) $ measurements for different Mn$ _3 $Sn samples provide an excellent correlation between the magnetic structures and the observed AHE. To further establish the connection, we have carried out isothermal magnetization measurements [$M(H)$] at different pressures at $ T= $ 100~K and 300~K, as shown in Fig.~5. The $ M(H) $ loops measured at 100~K  exhibit a nearly linear behavior up to 5~T and  remains almost unchanged with applied pressure up to 1.2~GPa [Fig 5(a)]. However, a close look to the low field regime [see the inset of Fig.~5(a)] reveals a unique pressure dependency of the magnetization. It is worth mentioning that the triangular AFM phase in Mn$ _3 $Sn displays a small spontaneous magnetization owing to a tiny deviation from the $120 ^{\circ}$ structure due to the alignment of Mn moments with their respective local easy axes \cite{localeasytilt}. In the present case, this residual spontaneous magnetization can even be seen for pressure of 0.6~GPa. When the system transforms to the helical AFM phase at higher pressure, the magnitude of the spontaneous magnetization starts decreasing. Finally, a nearly linear kind of $ M(H) $ loop is obtained for $ P= 1.2$~GPa. The disappearance of the residual spontaneous magnetization at higher pressures directly supports our previous finding of pressure induced helical magnetic state, where the uncompensated in-plane magnetic vector sums up to zero when it is rotated by $360^{\circ}$ in a single helical pitch. In corroboration with the 100~K Hall resistivity data, no pressure variation in the spontaneous magnetization is found for the $ M(H) $ loop measured at 300~K, as depicted in Fig.~5(b).  These findings thoroughly establish the correlation between the pressure induced switching of the AHE and magnetic ground states.

 
 Finally, pressure dependent XRD measurement is carried out at room temperature to find out the role of pressure on the crystal structure. Figure~6(a) shows the variation of lattice parameters $ a $ and $ c $ with hydrostatic pressure for Mn$_{3.05}$Sn$_{0.95}$. Both $ a $ and $ c $ decrease linearly with increasing pressure without showing any signature of structural phase transition up to a pressure of 10~GPa (see the XRD pattern in Appendix B). The unit cell volume ($ V $) also changes linearly with pressure [see the inset of Fig.~6(a)]. Therefore, we rule out any role of structural phase transition in the observed switching of AHE. We have also performed temperature dependent XRD study at ambient pressure for Mn$_{3.05}$Sn$_{0.95}$ to find out any structural anomaly at low temperatures [Fig.~6(b)]. Both $ a $ and $ c $ as well as $ V $ [inset of Fig. 6(b)] show a monotonic decrease with decreasing temperature, ruling out any temperature dependence of structural phase transition. It is also important to find out the relation between the crystal and magnetic structure in the helical AFM phase. Since our XRD measurements under pressure are limited to room temperature, we have selected Mn$_{3.03}$Sn$_{0.97}$ for the temperature dependent XRD study as this sample exhibits ambient pressure helical phase transition around 220~K. The temperature variation of lattice parameters for Mn$_{3.03}$Sn$_{0.97}$ are shown in Fig.~6(c). Interestingly, the in-plane lattice parameter $ a $ and $ V $ display a dip-type feature at the helical phase transition. It is to be noted here that our theoretical study emphasizes the role of pressure induced  trimerization of Mn atoms, where the in-plane bond length between neighbouring Mn triangles changes, in stabilizing the helical AFM phase. Hence, the observed anomaly in the temperature dependent $a$ and $V$ [inset of Fig.~6(c)] with almost no change in the out-of-plane lattice parameter $c$ corroborate our theoretical proposition.
 
  Although hydrostatic pressure stabilizes the helical phase at low temperature for Mn$_{3.05}$Sn$_{0.95}$, a similar transition is also found at ambient pressure for Mn$_{3.03}$Sn$_{0.97}$ with a larger unit cell volume. Here we note that irrespective of its exact composition, Mn$_3 $Sn intrinsically exhibits some amount of trimerization of the Mn atoms even at ambient pressure \cite{ANEmn3sn}, consistent with our theoretical calculations. The stability of the helical phase depends on the relative strength of $J_3$ in the presence of competing $J_2^l$ and $J_2^s$ exchange constants. In the ideal structure of Mn$ _3 $Sn, there are six nearest neighbours for the Mn-Mn exchange constants connected by $J_2$ [three above and three below, see Fig 1(b)]. The trimerization process splits the $J_2$ into a group of four $J_2^l$ and two $J_2$ [see Fig. 1(c)]. Hence, an overall decrease in the strength of $J_2$ is expected upon trimerization. As we can see from the phase diagram presented in Fig \ref{S2} (Appendix A), the $J_3$ value is very sensitive in stabilizing the helically modulated phase. It is also evident from Fig. 1(d) that an overall decrease in the strength of $J_2$ increases the helical phase region. Therefore, the helically modulated phase is a result of the subtle balance between $J_3$ and the splitting of $J_2$ due to the trimerization. The higher cell volume in case of  Mn$_{3.03}$Sn$_{0.97}$ may intrinsically give lower effective $J_2$ values in comparison to that of Mn$_{3.05}$Sn$_{0.95}$. In addition, different concentration of Mn atoms in these two samples also modifies the exchange interactions in the system. Hence, the observed different pressure effects in Mn$_{3.05}$Sn$_{0.95}$ and  Mn$_{3.03}$Sn$_{0.97}$ depends not only on the volume of the unit cell, also on the position of extra Sn and the effective exchange interaction strength in the system. Therefore, both positive and negative pressure may cause the helical AFM transition.

The present study exemplifies the mechanism for the switching of AHE in Mn$_3$Sn system by establishing a route to control the modulation by means of pressure. As the exchange interaction between the Mn moments depends directly on the bond length, the switching of ground state by exchange parameter variation can be achieved by control over the lattice parameters by means of pressure/strain. Though chemical pressure can bring about the required change in the lattice parameters but it cannot be used for the switching purpose in device applications. Hydrostatic pressure, on the other hand, has been widely used to manipulate the lattice parameters and even to bring about a structural transition in a system.  The present mechanism to manipulate AHE differs completely from the recent reports of AHE switching in antiferromagnets \cite{natelecSwitch,mn3gePressureSwitch,mn3snEswitch_nakatsuji}. Switching of AHE in thin films of Mn$_3$Pt relies on the structural distortion controlled through electrical response in a piezo-electrical substrate, whereas, in case of bulk Mn$_3$Ge the underlying mechanism for the pressure induced sign change of AHE is still unknown. In the present work, we have uncovered the process of trimerization in Mn$_3$Sn that was not taken into consideration till now. In such a scenario, three nearest-neighbour Mn moments collectively respond to any external stimuli and in turn can greatly affect the magnetic properties at higher trimerization levels \cite{gd3trimer,Gd3helicaltrimer}. Now that we understand the ingredients for the stabilization of different ground states, the effect of Mn doping can also be explored to manipulate exchange parameters that will help to achieve switching at room temperature. 

\section{CONCLUSION}

In conclusion, we have successfully demonstrated the switching of anomalous Hall effect in antiferromagnetic Mn$_3$Sn using hydrostatic pressure. The theoretical analyses presented in this work establish a first hand link between different magnetic ground states of Mn$_3$Sn.  Our experimental results convincingly manifest the role of helical modulation of the inverse triangular spin structure to manipulate the AHE in the system. The pressure induced  effective trimerization of the Mn atoms controls the exchange frustration which finally leads to the helical modulation. The present  work could immensely help in the greater understanding of extensive range of properties demonstrated by Mn$ _3 $Sn and related materials.

A.K.Nayak acknowledges the support from Department of Atomic Energy (DAE), the Department of Science and Technology (DST)-Ramanujan research grant (Grant No. SB/S2/RJN-081/2016), SERB research grant (Grant No. ECR/2017/000854) and Nanomission research grant [Grant No. SR/NM/NS-1036/2017(G)] of the Government of India. A.K.Nandy acknowledges the support of DAE, SERB research grant (Grant No. SRG/2019/000867) of the Government of India. A.K.Nandy thanks Prof. P. M. Oppeneer for the Swedish National Infrastructure for Computing (SNIC) facility and Dr. Gustav Bihlmayer for the computational support, JURECA at JSC, Forschungszentrum, J\"ulich, Germany. VS and RN acknowledge BRNS, India, for financial support bearing sanction Grant No. 37(3)/14/26/2017.

\onecolumngrid
\vspace{20pt}
\hrule{}
\section{APPENDIX A: Theoretical Model}
\begin{figure}[h]
	\includegraphics[angle=0,width=12cm,clip]{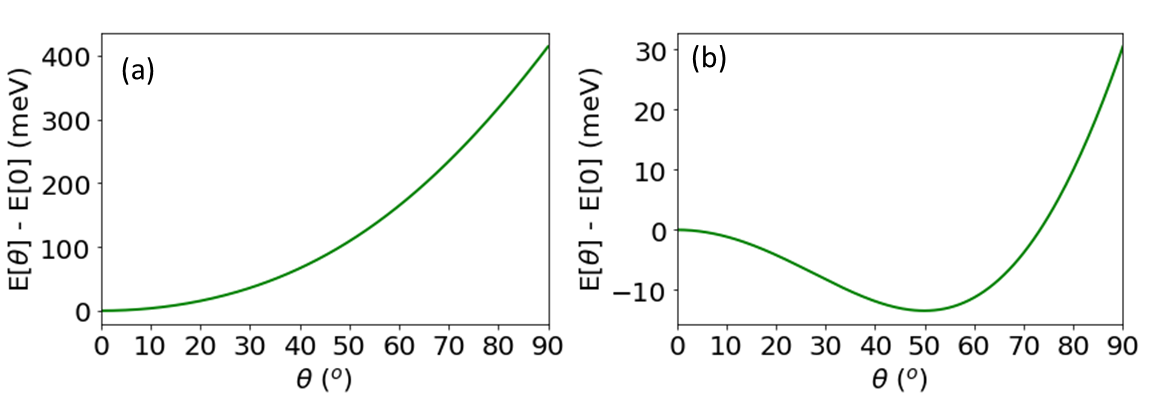}
	\caption{\label{S1}Two prototype plots as obtained for (a) exchange parameters supporting a non-modulated structure and (b) exchange parameters supporting a modulated structure. }	
\end{figure}

To understand the role of competing exchange interaction in stabilizing the helical modulation, we write down the classical Heisenberg Hamiltonian for a single Mn atom considering the inter Kagome sub lattice coupling J$_1$, J$_2$ and J$_3$ (see Eq. (\ref{Hemil})). As J$_1$ is the strongest among three exchange parameters \cite{mn3qm}, we varied J$_2$ and J$_3$ in the units of J$_1$ over a range to obtain the ground state for all combinations. For every particular combination, we change the value of the angle of rotation ($\theta$) between two layers from 0 to 90 degree in order to determine the lowest energy state. After performing such calculations, two prototypical plots are shown in Fig \ref{S1}. The value of $\theta$ corresponding to the energy minima for these plots is assigned as the magnetic ground state. Here, Fig \ref{S1} (a) represents an unmodulated inverse triangular ground state while Fig \ref{S1} (b) shows the spin-spiral (SS) energy for a helically modulated ground state. Similar calculations were performed for all the combinations and the corresponding phase diagrams are constructed. 

\begin{figure}[h]
	\includegraphics[angle=0,width=13cm,clip]{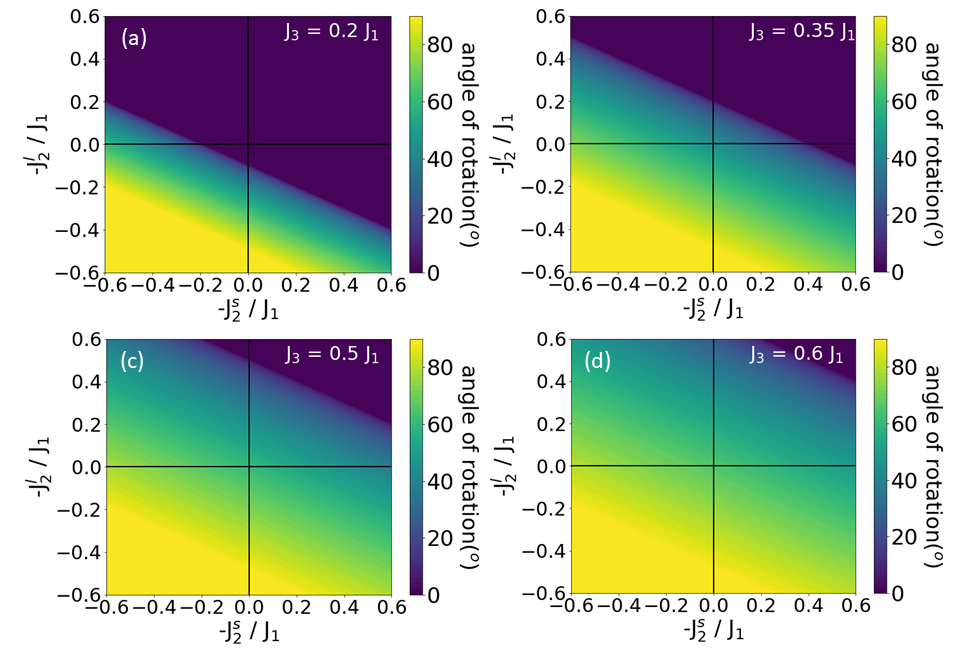}
	\caption{\label{S2}Phase diagrams for the angle of rotation between two adjacent Kagome layers in the J$_2^s$-J$_2^l$ plane with J$_3$ fixed at (a) 0.2 J$_1$, (b) 0.35 J$_1$, (c) 0.5 J$_1$ and (d) 0.6 J$_1$. }	
\end{figure}

In case of trimerization with unequal equilateral triangles in the Kagome sublattices, the bond corresponding to J$_2$ now splits into two unequal lengths. The Hamiltonian for this case is written as

\begin{multline} \label{Hsub}
H = 2J_1\cos(\frac{2\pi}{3}+\theta) +2J_1\cos(\frac{2\pi}{3}-\theta) +2J_2^l\cos(0+\theta) +2J_2^l\cos(0-\theta) +J_2^s\cos(0+\theta) +J_2^s\cos(0-\theta) \\
+J_3\cos(0+2\theta) +J_3\cos(0-2\theta)
\end{multline}
where, J$_2^l$ (J$_2^s$) is the exchange constant corresponding to the larger (smaller) bond length. Similar to the previous calculations, here, we have performed the energy calculations for all sign and strength combinations of J$_2^l$ and J$_2^s$ where J$_1$ and J$_3$ are fixed at constant values. Main text includes the phase diagram in the J$_2$-J$_3$ plane measured in the unit of J$_1$. In case of effective trimerization, Fig \ref{S2} contains the phase diagrams for various relative values of J$_3$ ranging from J$_3$/J$_1$ = 0.2 to 0.6. From Fig \ref{S2}, we conclude that the helical ground state can be achieved for various relative values of J$_3$. Importantly, the area of the SS phase in the J$_2^l$ -J$_2^s$ plane is increasing with the relative strength of J$_3$. Here, we mention that these values of J$_3$ are from different regime of the phase diagram calculated for ideal Kagome structure (marked in Fig 1 (d)). J$_3$ = 0.2, 0.35, 0.6 are taken outside the helical phase boundary, close to the phase boundary and deep inside the helical phase, respectively. We also note that the sign change of either J$_2^l$ or J$_2^s$ favors the helically modified ground state, supporting our exchange constant values extracted from the DFT calculations. An increase in J$_3$ moves the phase boundary between zero and finite helical modulation towards higher J$_2$ and vice versa (Figure \ref{S2}).

\subsection{Spin-spiral calculations}
\begin{figure}[h]
	\includegraphics[angle=0,width=17.5cm,clip]{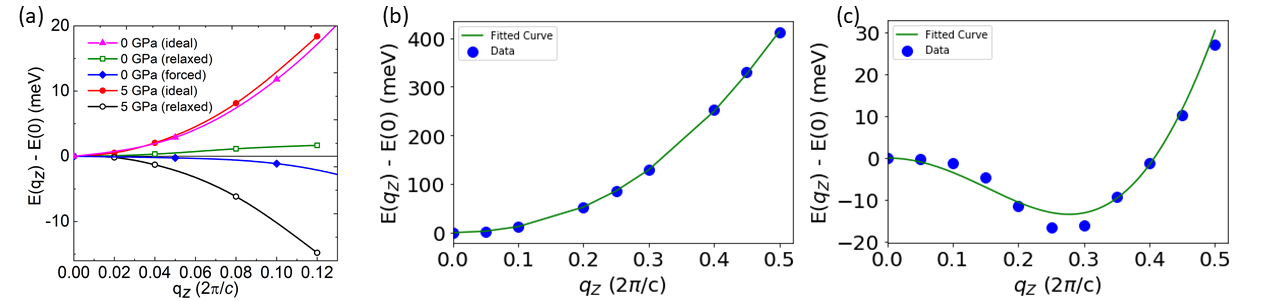}
	\caption{\label{S3}(a) Spin spiral energy $E({\bf q} )$ vs ${\bf q} (00q_z)$ as calculated for various pressure values with and without effective trimerization. $E(q_z)$ vs $q_z$ data points along with fitted curves for (b) Ideal and (c) forced trimerize Kagome sublattice at ambient pressure. }	
\end{figure}

The SS calculations were performed for different level of trimerization due to nonzero pressure in the system. Fig. \ref{S3} (a) shows the energy as a function of spin spiral vector $\textbf{q}$ $=$ $(0 0 q_z)$ plot close to {\bf q} $=$ 0 (main text includes the plot for full range of {\bf q}). Note, the above $\theta$ value can be translated to $q_z$ in the SS calculations. From Fig \ref{S3} (a), we clearly see that in case of ideal Kagome sublattice stacking, the magnetic state does not show any helical modulation even at high pressure till 5 GPa. After structure relaxation, the energy curve at ambient pressure gets softer than that for ideal structure, but {\bf q} $=$ 0 still remains the magnetic ground state. However, the softening of the dispersion displays the importance of effective trimerization which is originating from the size mismatch of Mn and Sn atoms i.e. the chemical pressure effect. Further increase in the trimerization at ambient pressure ($\delta$ $=$ 0.12) leads to a helical ground state corresponding to the $q_z$ $\neq$ 0 state. For the relaxed structure at 5 GPa pressure, where the effects of trimerization is higher ($\delta$ $=$ 0.103) than that at ambient pressure ($\delta$ $=$ 0.024), the energy curve shows a large negative slope. This establishes a helically modulated SS to be the ground state with applying hydrostatic pressure as our experimental observations suggest. The SS curves were fitted using the Hamiltonian as in Eq. (\ref{Hemil}) for ideal case, and Eq. (\ref{Hsub}) for the trimerized case. Fitted profiles are shown in Fig. \ref{S3} (b-c). The fitted exchange parameters are tabulated in Table-\ref{tab2}.

\begin{table}[h]
	\caption{\label{tab2} Out-of-pane exchange parameters as extracted by fitting the $E({\bf q})$ vs $q$ curves with the usual exchange Hamiltonian.}
	\begin{tabular}{|l|l|l|l|l|}
		\hline
		\multirow{2}{*}{} & J$_1$ (meV) & \multicolumn{2}{l|}{J$_2$ (meV) {[}-J$_2$/J$_1${]}}                                                   & \multirow{2}{*}{J$_3$ (meV) {[}J$_3$/J$_1${]}} \\ \cline{2-4}
		&          & \begin{tabular}[c]{@{}l@{}}J$_2^l$ (meV) {[}-J$_2^l$/\\ J$_1${]}\end{tabular} & J$_2^s$ (meV) {[}-J$_2^s$/J$_1${]} &                                       \\ \hline
		0 GPa (Ideal)     & 20.83    & \multicolumn{2}{l|}{-9.29 {[}0.445{]}}                                                       & 7.09 {[}0.340{]}                      \\ \hline
		0 GPa (Forced)    & 13.12    & -0.32 {[}0.024{]}                                                  & 2.59 {[}-0.197{]}       & 4.41 {[}0.336{]}                      \\ \hline
	\end{tabular}
\end{table}

\newpage

\section{APPENDIX B: X-ray Diffraction (XRD) Measurements}
\begin{figure}[h]
	\includegraphics[angle=0,width=16cm,clip]{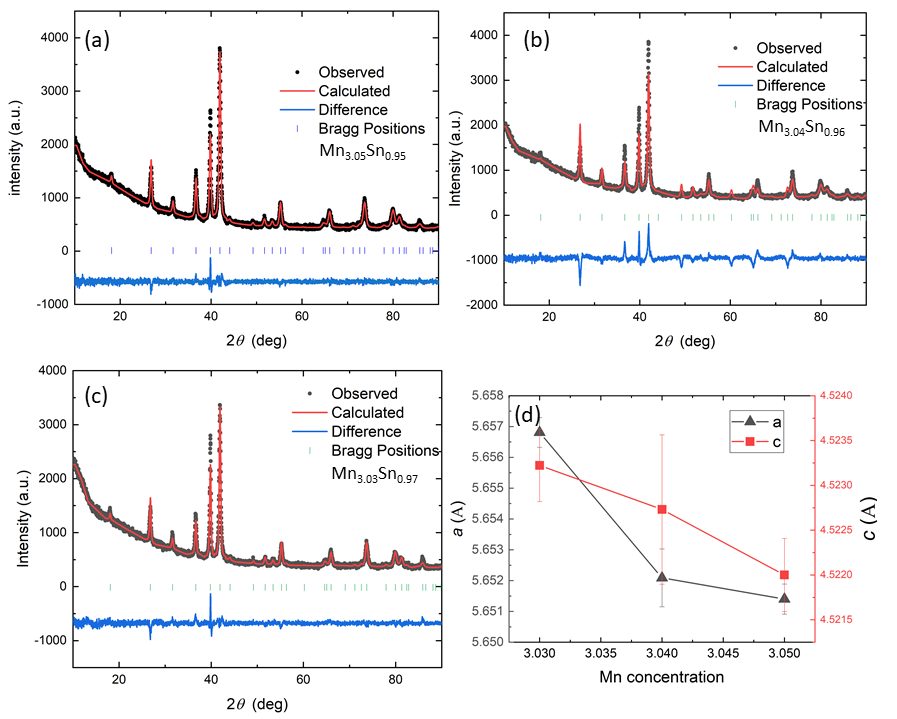}
	\caption{\label{S4}XRD pattern with Rietveld refinement for (a) Mn$_{3.05}$Sn$_{0.95}$, (b) Mn$_{3.04}$Sn$_{0.96}$, and (c) Mn$_{3.03}$Sn$_{0.97}$ sample. (d) Calculated $a$ and $c$ values from the fitting. }	
\end{figure}

XRD measurements were performed on polycrystalline samples using a Rigaku X-ray diffractometer. Figure \ref{S4}(a)-(c) shows the fitted profile for different composition of Mn$_{3+x}$Sn$_{1-x}$ samples. Fittings were performed taking into care the extra Mn concentration. Occupancy of respective sites were modified according to extra Mn. No extra impurity peaks were observed for any sample. Refined lattice parameters are plotted in Fig. \ref{S4}(d). We find a small monotonic change in the lattice parameters with Mn composition.

\subsection{Pressure dependent XRD measurements}
\begin{figure}[h]
	\includegraphics[angle=0,width=16cm,clip]{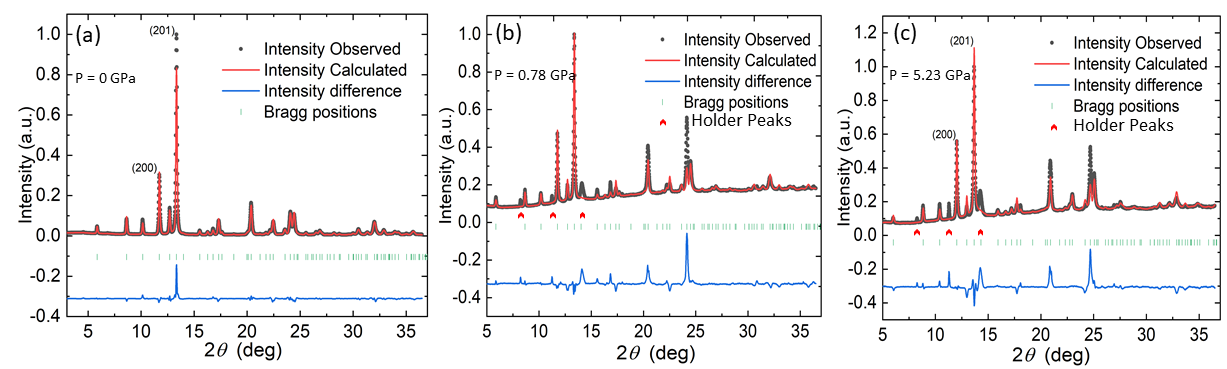}
	\caption{\label{S5} XRD patterns with Rietveld refinement measured at (a) 0 GPa, (b) 0.78 GPa, and (c) 5.23 GPa. }	
\end{figure}

Pressure dependence of XRD measurements were performed using ECXRD (BL-11) beamline at synchrotron radiation source beam facility at RRCAT, Indore, India. Figure \ref{S5} (a) shows the ambient pressure XRD pattern with fitted profile for Mn$_{3.05}$Sn$_{0.95}$ sample. XRD pattern performed at such high fluence again establishes the single phase nature of sample. Fig. \ref{S5} (b)-(c) show the fitted profile at finite pressure values. The presence of some additional peaks marked inside the figure originate from the high pressure gasket material used in diamond anvil cell. We find that with the pressure cell in place, the intensity mismatch between the experimental and simulated patterns increases for some reflections. This may be due to some preferred orientation of grains inside the pressure cell. This mismatch also explains the presence of larger error bars in lattice parameters obtained in presence of pressure in the main text.

\newpage

\subsection{Low temperature XRD measurements}
\begin{figure}[h]
	\includegraphics[angle=0,width=15cm,clip]{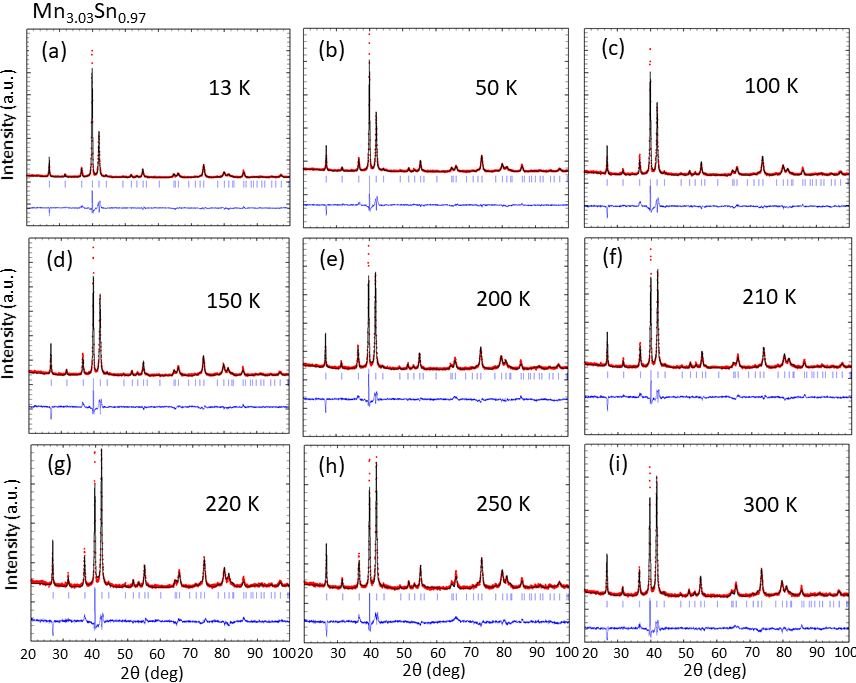}
	\caption{\label{S6}  Profile fitting graphs of XRD for Mn$_{3.03}$Sn$_{0.97}$ sample at various temperatures. Red circles represent the experimental data. Black line shows the calculated fitting. Difference between calculated and experimental graphs is shown as blue line. Straight blue ticks mark various Bragg’s positions. }	
\end{figure}

\begin{figure}[h]
	\includegraphics[angle=0,width=16cm,clip]{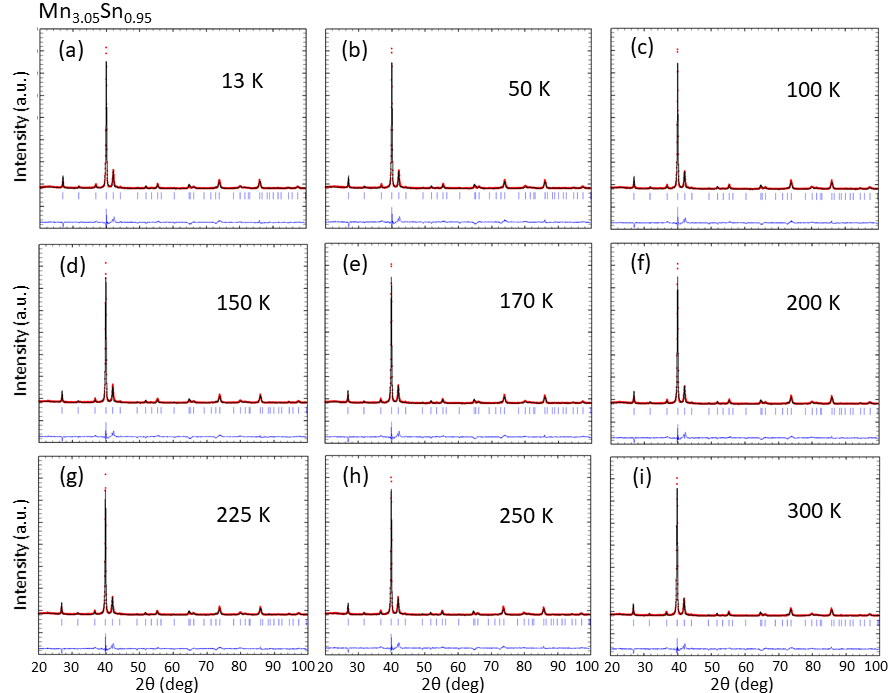}
	\caption{\label{S7} Profile fitting graphs of XRD for Mn$_{3.05}$Sn$_{0.95}$ sample at various temperatures. Red circles represent the experimental data. Black line shows the calculated fitting. Difference between calculated and experimental graphs is shown as blue line. Straight blue ticks mark various Bragg’s positions. }	
\end{figure}

Low temperature XRD measurements were performed using a low temperature (Oxford Phenix) attachment to the X-Ray diffractometer (PANalytical). Fig. \ref{S6} and Fig. \ref{S7} show the XRD patterns with fitting at different temperatures. No structural transition is visible down to 13 K in both the samples. Lattice parameters were first refined for room temperature XRD. Then, the refined parameters were used as a starting point for the refinement of the subsequent lower temperature. Similar process was followed for all the temperatures. Fitted lattice constants $a$ and $c$ are presented in the main text.

\newpage

\section{APPENDIX C: Resistivity measurements}
\begin{figure}[h]
	\includegraphics[angle=0,width=12cm,clip]{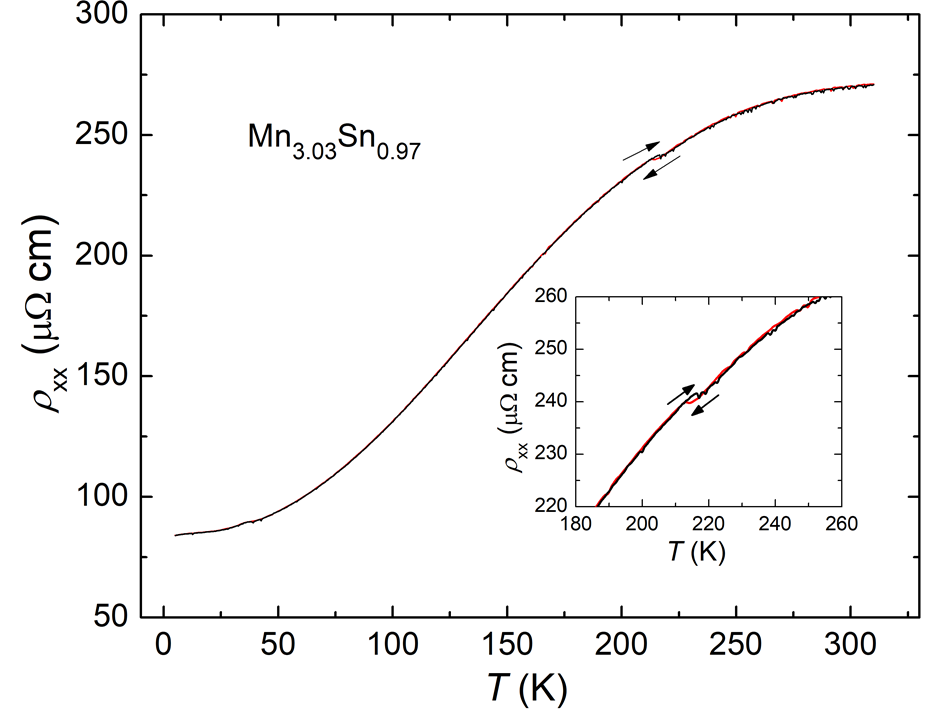}
	\caption{\label{S8} Temperature dependence of resistivity measured at ambient pressure for Mn$_{3.03}$Sn$_{0.97}$. The inset shows the amplified view of the resistivity data around 200 K. The presence of hysteresis between the cooling and heating resistivity data signifies the presence of a first-order phase transition. }	
\end{figure}

Temperature dependence of resistivity measured at ambient pressure for Mn$_{3.03}$Sn$_{0.97}$ is shown in Fig. \ref{S8}.

\newpage
\twocolumngrid

\end{document}